\documentclass[aps,prl,superscriptaddress,preprint,amsmath,amssymb,graphicx]{revtex4}

\newif\ifstructure
\structuretrue

\newcommand{\beq}{\begin{equation}}
\newcommand{\eeq}{\end{equation}}
\newcommand{\bea}{\begin{eqnarray}}
\newcommand{\eea}{\end{eqnarray}}
\providecommand{\abs}[1]{\left\lvert#1\right\rvert}

\providecommand{\bra}[1]{\langle #1 \rvert}
\providecommand{\ket}[1]{\lvert #1 \rangle}

\usepackage{afterpage}
\usepackage{empheq}
\newcommand{\Heff}{H_{\text{eff}}}

\usepackage{subcaption}
\usepackage[utf8]{inputenc}

\usepackage{tikz}
\usepgflibrary{arrows}
\usetikzlibrary{decorations}
\usetikzlibrary{decorations.pathmorphing,patterns}
\usetikzlibrary{scopes}
\usetikzlibrary{shapes, matrix}

\usepackage{tikz}
\usepgflibrary{arrows}
\usetikzlibrary{decorations}
\usetikzlibrary{decorations.pathmorphing,patterns}
\usetikzlibrary{scopes}
\usetikzlibrary{shapes, matrix}
\usepackage{xcolor}

\begin{document}

\title{Non-Hermitian Hamiltonians for Linear and Nonlinear Optical Response: a Model for Plexcitons}

\author{Daniel Finkelstein-Shapiro \footnote{daniel.finkelstein@iquimica.unam.mx} }
\affiliation{Instituto de Qu\'{i}mica, Universidad Nacional Aut\'{o}noma de M\'{e}xico, CDMX, M\'{e}xico}
\author{Pierre-Adrien Mante}
\affiliation{Division of Chemical Physics and Nanolund, Lund University, Box 124,
221 00 Lund, Sweden}
\author{Sinan Balci}
\affiliation{Department of Photonics, Izmir Institute of Technology, 35430 Izmir,
Turkey}
\author{Donatas Zigmantas}
\affiliation{Division of Chemical Physics and Nanolund, Lund University, Box 124,
221 00 Lund, Sweden}
\author{T\~{o}nu Pullerits 
\footnote{tonu.pullerits@chemphys.lu.se} }
\affiliation{Division of Chemical Physics and Nanolund, Lund University, Box 124,
221 00 Lund, Sweden}

\begin{abstract}
In polaritons, the properties of matter are modified by mixing the molecular transitions with light modes inside a cavity. Resultant hybrid light-matter states exhibit energy level shifts, are delocalized over many molecular units and have a different excited-state potential energy landscape which leads to modified exciton dynamics. 
Previously, non-Hermitian Hamiltonians have been derived to describe the excited states of molecules coupled to surface plasmons (i.e. plexcitons), and these operators have been successfully used in the description of linear and third order optical response.
In this article, we rigorously derive non-Hermitian Hamiltonians in the response function formalism of nonlinear spectroscopy by means of Feshbach operators, and apply them to explore spectroscopic signatures of plexcitons. In particular we analyze the optical response below and above the exceptional point that arises for matching transition energies for plasmon and molecular components, and study their decomposition using double-sided Feynman diagrams. 
We find a clear distinction between interference and Rabi splitting in linear spectroscopy, and a qualitative change in the symmetry of the lineshape of the nonlinear signal when crossing the exceptional. This change corresponds to one  in the symmetry of the eigenvalues of the Hamiltonian.  
Our work presents an approach for simulating the optical response of sublevels within an electronic system, and opens new applications of nonlinear spectroscopy to examine the different regimes of the spectrum of non-Hermitian Hamiltonians. 
\end{abstract}

\maketitle

\section{Introduction}


The coupling between light modes and radiative transitions of matter creates exciton polariton states, i.e. half-light and half-matter states. These were first demonstrated in atoms inside a microwave cavity and were used to study decoherence and quantum logic operations in atoms \cite{Walther2006,Mabuchi2002,Haroche1989,Raimond2001}. More recently, polaritonic states have also been observed in molecules in microcavities, having both infrared and visible wavelengths. 
They open new paradigms with which to influence chemical processes \cite{Kasprzak2006a,Martinez-Martinez2018,Yuen-Zhou2016,Zhong2016,Feist2015,Schachenmayer2015,Herrera2016,Galego2015,Hutchison2013,Hutchison2012,Thomas2016,Wang2014a}. For example, the delocalization of polaritonic states is advantageous for long-distance energy transfer \cite{Coles2014,Saez2018,Xiang2020,
Du2018,Zhong2017,Krainova2020,Herrera2016}, polaritonic chemistry \cite{Du2019,Fregoni2020,Ribeiro2018,Feist2018} and photocatalysis \cite{Mandal2019,Manuel2019}. They can also be used for lasing and to study nonequilibrium condensates and phase transitions \cite{Kasprzak2006,Keeling2020}. 
Using plasmonic nanoparticles instead of cavities confines the light to subwavelength dimensions. This allows creating billions of polaritonic systems in solution \cite{Balci2014}.  It allows many more interesting arrangements that exhibit strong coupling in the single or few molecule limit \cite{Chikkaraddy2016}.

The number of molecules in the interaction volume and the properties of the cavity determine the light-matter coupling strength, while the quality of the cavity, and the molecular dissipative processes determine the coherence lifetime. The coupling strength and coherence lifetimes span a large parameter space that can be broken into different regimes, which are also associated to different phenomena \cite{Torma2015,Pelton2019}. For example, for weak coupling we observe a Purcell enhancement of the fluorescence rate while for strong coupling we see a Rabi splitting, which is reflected in a splitting of the absorption into two peaks termed upper and lower polariton branches. However, such an apparent splitting is also possible in the presence of an interference that appears already in the weak coupling regime (also called electromagnetically induced transparency, or a Fano interference \cite{Pelton2019}). Finally, the ultrastrong coupling and deep strong coupling regimes are associated to the breaking of the rotating-wave approximation and are difficult to reach using molecules in cavities, but have been observed with plasmonic lattices \cite{Mueller2020}.

It is important to discuss dissipation when considering polariton transitions. While in atoms the sources of dissipation are few and well understood, an intense experimental and theoretical effort has been necessary to assign the timescale and processes of coherent and incoherent dynamics in molecular polaritons \cite{Fofang2011,Balci2014,Hranisavljevic2002,Vasa2013,Wiederrecht2008,Hoang2016,Finkelstein2021}. 
Briefly, time-resolved spectroscopy experiments have detected coherent dynamics in the form of Rabi oscillations via transient absorption \cite{Vasa2013}, and established the possible relaxation pathways: the upper branch can decay via radiation damping to the ground state \cite{Virgili2011}, inject into the dark plexciton states \cite{DelPo2020,Xiang2019}, or decay directly into the lower branch \cite{Mewes2020}. The dark states themselves can decay to the lower polariton branch, with some degree of back-transfer (from lower polariton to dark reservoir) taking place \cite{Luk2017}. The long lifetime of the dark states is responsible for the excited state stabilization with decay times for up to several $\mu s$ \cite{Xiang2018,Zhang2019,DelPo2020}.  
In systems, where the cavity mode is confined to a plasmonic nanoparticle, the dynamics are strongly influenced by electron-electron and electron-phonon scattering inside the metal nanoparticle. Notably, a direct coupling between the molecular dark states and metal surface states appears to limit the lifetime of the dark states to that of the electron-electron scattering time \cite{Finkelstein2021}.  

Simulations of the optical response are crucial to correctly interpret the experiments, and more so in time-resolved spectroscopy experiments such as two-dimensional coherent spectroscopy (2DCS) or transient absorption. Coupled oscillator models work well to understand linear optical absorption \cite{Mewes2020}. For transient absorption they can also work as long as the experiment can be simulated as two linear experiments, the bleach of the ground state, and the absorption of an excited state dominated by the signal from the remaining molecules in the ground state. This is true when the main nonlinearity is a Rabi contraction whereby the Rabi splitting of the ground state is reduced because of the molecules that have been excited  \cite{Vasa2013}. However, they are not sufficient for simulating a third-order signal as in general all orders of the field appear together and it is not straightforward to disentangle them. Input-output theory has also been used to compute with a very good agreement the experimental two-dimensional infrared spectroscopy (2DIR) signal of molecules in infrared cavities \cite{Ribeiro2018}. The Rabi contraction and softening of the vibrational mode (for 2DIR) have been proposed as the mechanisms of nonlinearity. Similar physics has been found in polaritons formed from molecules adsorbed on plasmonic lattices \cite{Vasa2013}. In an earlier work, we have simulated the two-dimensional electronic spectroscopy response of molecules coupled to plasmonic nanoparticles using an extension of the response function formalism to non-diagonal non-Hermitian Hamiltonians, successfully reproducing the spectra at early times using a Rabi contraction nonlinearity and at late times using a thermal expansion of the nanoparticle \cite{Finkelstein2021}. Gu et al. have also employed a non-Hermitian Hamiltonian to calculate several nonlinear experiments, finding significant modifications of the energy levels as well as the selection rules \cite{Gu2021}. Models of disordered polaritons have suggested the existence of exceptional points in these non-Hermitian Hamiltonians which vary according to the degree of disorder \cite{Engelhardt2022} and these have been measured experimentally in infrared cavities \cite{Cohn2022}. 
Non-Hemiticity of the Hamiltonian can arise from a number of different scenarios, not only due to energy relaxation but also from particle decay for example in photoionization. Moiseyev et al. has proposed a more realistic model of molecules including the ionized continuum states, and studied their modification inside a cavity \cite{Moiseyev2022}. 

There are two difficulties related to the description of the dynamics of hybrid plasmonic systems. First, the description of the optical response in terms of non-diagonal non-Hermitian Hamiltonians is not typical in the approach for calculating 2D spectra in molecular systems. Usually an excitonic basis is used, where diagonal fluctuations induced by the bath are added and result in a lineshape function after bath-averaging \cite{Cho2009}. Second, plasmonic systems add a rich dimension of dynamics in terms of the non-equilibrium distribution of hot electron-hole pairs, formed inside the metallic band structure after plasmon decay. The dynamics inside the metallic band observed in transient absorption experiments has been beautifully demonstrated in pure metals \cite{Brown2017}, however the description has not yet been extended to mixed molecule-metal systems. One complication is the co-existence of coherent effects related to the Rabi oscillation between two discrete transitions along with the incoherent scattering of continuous non-equilibrium distribution of electrons in the metal. The latter phenomena is described by the dielectric constant of an electron gas whose temperature is described by the two-temperature model \cite{Voisin2001} and has not been adapted to handle the molecular states.

In this work, we consider the linear and nonlinear optical response of non-diagonal non-Hermitian Hamiltonians. We begin by providing a  derivation of the  non-Hermitian Hamiltonian and discuss its regime of validity. We then describe the decomposition of the linear and third-order optical response using double-sided Feynman diagrams. We conclude with a discussion of the polariton branch energy structure in the weak and strong coupling regimes, and their distinctive spectral signatures.         

\section{Theoretical description}

\textbf{Effective operators.} Non-Hermitian Hamiltonians have been derived as effective operators for molecules coupled to plasmonic nanoparticles from  \textit{ab initio} considerations \cite{Delga2014}. The non-Hermiticity arises from an implicit inclusion of certain degrees of freedom of the system. Regardless of the procedure to obtain the effective operators, one is left with a $2 \times 2$ Hamiltonian describing a plasmon excitation coupled to a molecular excitation where each of these two transitions is associated to a separate decay channel. In the site basis, consisting of the molecular bright state and the plasmon transition, the Hamiltonian is written as (setting $\hbar=1$)
\begin{equation}
    \Heff = \begin{bmatrix}
    \omega_J-i\gamma_J & g \\
    g & \omega_P-i\gamma_P
    \end{bmatrix}
    \label{eq:Heff}
\end{equation}
where $g$ is the coupling between the bright molecular transition and the plasmon transition, the $P$ and $J$ indices denote plasmon or molecular J-aggregate, and $\omega_i$, $\gamma_i$ relate to the transition frequencies and dephasing rates, respectively. We impose $\gamma_J<\gamma_P$. The Hamiltonian of equation \eqref{eq:Heff} admits an exceptional point when $\omega_J=\omega_P\equiv \omega_0$ and $g=\frac{\Delta \gamma}{2}$ where $\Delta \gamma = \gamma_P-\gamma_J$ \cite{Rodriguez2016}, and which sets the limit between weak and strong coupling. 

We can easily solve for the eigenvalues $\omega_{\pm}$ and eigenvectors $\vec{v}_{\pm}$ of the Hamiltonian as:
\begin{equation}
    \omega_{\pm} =  \omega_0-i\frac{\gamma_J+\gamma_P}{2} \pm \sqrt{g^2-(\Delta\gamma/2)^2}  \equiv \omega_0-i\bar{\gamma} \pm \xi
    \label{eq:eigenvalues}
\end{equation}
 and
\begin{equation}
    \vec{v}_{\pm} = \frac{1}{\sqrt{(\pm \xi+i\Delta \gamma/2)^2+g^2}}\begin{bmatrix}
    \pm \xi+i\Delta \gamma/2 \\
    g
    \end{bmatrix}
    \label{eq:eigenvectors}
\end{equation}
where we will denote by $+$ and $-$ the upper and lower polariton branches, respectively. The transition dipole moments are calculated as $\mu_{\pm g} = \vec{v}_{\pm}^T \begin{bmatrix} \mu_J \\ \mu_P \end{bmatrix}$ where $\mu_P$ and $\mu_J$ are the transition dipole moments of the plasmon and bright molecular transition, respectively. We assume that the coupling $g$ is real. $\xi$ can be real or imaginary and the transition from one to the other appears at the exceptional point. Given that the matrix in Eq. \eqref{eq:effective_Hamiltonian} is symmetric, we have $\mu_{\pm g}=\mu_{g \pm }\equiv \mu_{\pm}$. 
For $\mu_{P}\neq 0, \mu_J = 0$, we have $       \mu_{\pm} = \mu_P\frac{g}{\sqrt{(\pm \xi+i\Delta \gamma/2)^2+g^2}}$. The denominator can be rewritten as $g\sqrt{(-i\delta \pm \sqrt{1-\delta^2})^2+1}$ where $\delta=\frac{\Delta\gamma}{2g} $. Doing a perturbative expansion on either side of the exceptional point $\delta = 1 \pm \epsilon$ for small $\epsilon$, we have to leading order $\mu_{\pm} \approx \sqrt{\pm\sqrt{2\epsilon}}$ in the weak coupling regime, making one transition dipole moment purely real, and the other purely imaginary. 
In the strong coupling limit, to leading order we have $\mu_{\pm} \approx \sqrt{2\epsilon \mp i \sqrt{2\epsilon}}$ which leads to the relation $\mu_+^2=(\mu_-^2)^*$. These relations will be important when deriving the symmetries of the signals. The limits presented here, either $\mu_J =0$ or $\mu_P =0$ should be physically understood as  $\mu_J \ll \mu_P$ or $\mu_P \ll \mu_J$ since otherwise there would be no dipolar coupling between molecules and plasmons. 

The Hamiltonian from Eq. \eqref{eq:Heff} can be derived from classical arguments \cite{Rodriguez2016} or be obtained from the quantization of the modes of the electromagnetic field sustained by a spherical metallic particle \cite{Delga2014}. In Appendix A we show a derivation suitable for the parameters of plexcitons explored earlier \cite{Finkelstein2021} while in Appendix B we show a derivation obtained by projecting out continuum states.  However, the effective non-Hermitian operator remains an incomplete description for time-resolved experiments as it cannot capture the dynamics occurring in the implicit degrees of freedom, i.e. the degrees of freedom that when removed induce the non-Hermiticity. In our case these correspond to the excited electron-hole pairs in the metallic conduction band. The non-Hermitian can be used for calculating the linear response, but it is not justified for higher-order signals. The dephasing of the optical coherence can give a width to the lineshape, however when this width arises for instance from decay into a different excitation (i.e. electron-hole states), these will contribute to the excited-state absorption in a way that cannot be handled by the non-Hermiticity alone. We  provide a justification for its use under restricted conditions in the next section, and derive the expressions applicable to the general case. \newline

\textbf{Optical response.} The density matrix to n$-th$ order in the light-matter interaction $\rho^{(n)}(t)$ is given by (omitting the hat symbol for operators \cite{Mukamel1995}):
\begin{equation}
\begin{split}
    \rho^{(n)}(t) &= i^n \int dt_n \int dt_{n-1} \cdots \int dt_1 \\
    &\times  E(t-t_n) 
    E(t-t_n-t_{n-1}) 
    ... 
    E(t-t_n-...-t_1) \\
    &\times \langle [\mu(t_{n-1}+...+t_1),[\mu(t_{n-2}+...+t_1),...[\mu(0), \rho(-\infty)]] \rangle_{B}
    \label{eq:rho_n}
\end{split}
\end{equation}
where $\mu(t)=e^{+iH_0t}\mu e^{-iH_0t}$ is the transition dipole moment operators in the interaction picture, $H_0$ is the field-free Hamiltonian, $E(t)$ is the electric field from the impinging radiation and $\langle \rangle_{\text{B}}$ is the average over the realizations of the bath. 
We define the Feshbach operators $P$ and $Q$ for the bright and dark partitions, respectively. An illustration of the transformation from molecular and plasmonic partitions to bright and dark partitions is shown in Figure \ref{fig:Hamiltonian} (details are found in Appendix A).  

\begin{figure}
    \includegraphics[width=0.8\textwidth]{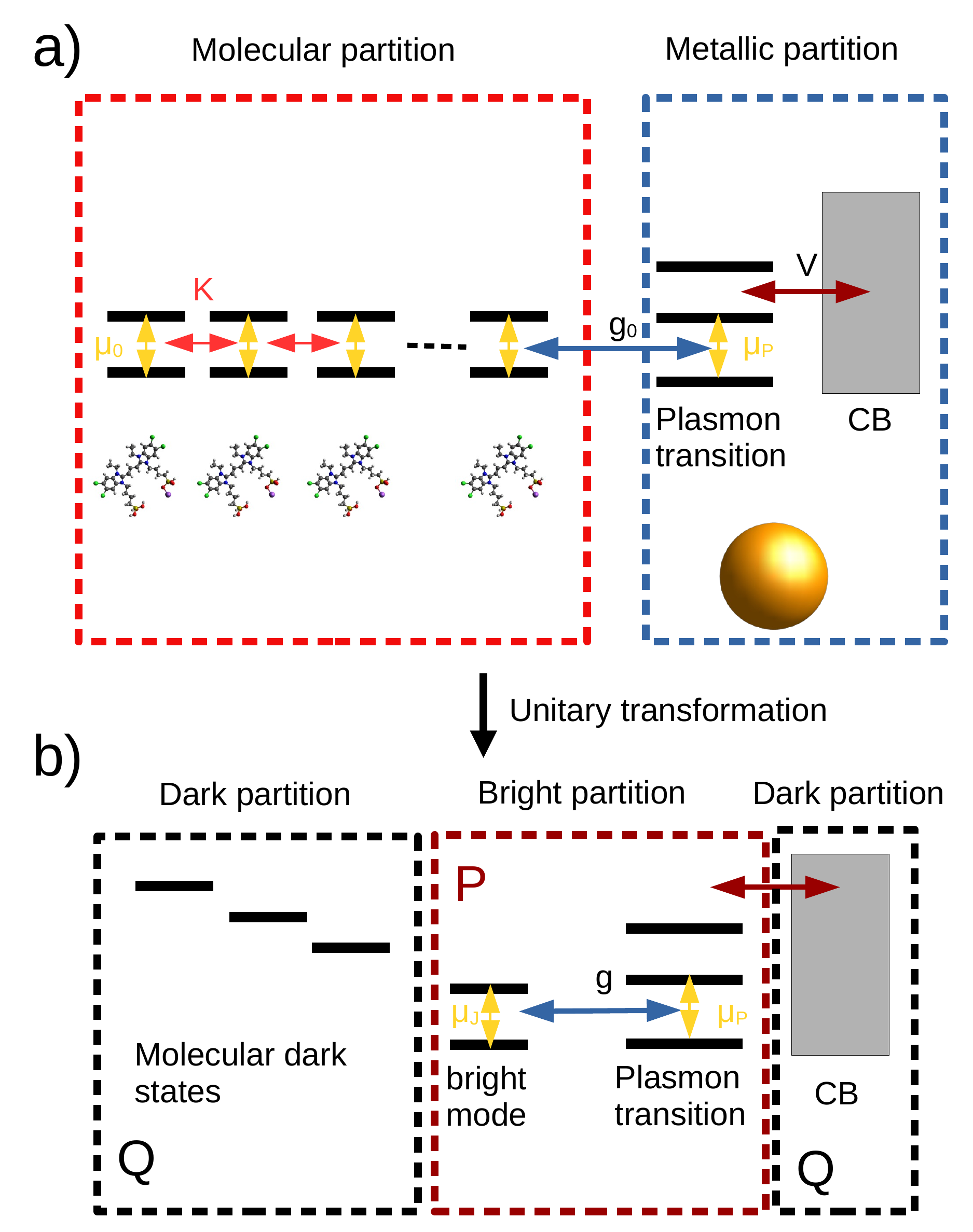}
    \caption{Energy level structure. a) Before diagonalization of the molecular component b) After diagonalization. Details are found in Appendix A}
    \label{fig:Hamiltonian}
\end{figure}

To remove the explicit description of the dark modes in Eq. \eqref{eq:rho_n}, we derive the elements needed to calculate the nested commutators in terms of effective operators. Because $P$ delimits the bright partition, the optical response function will be contained in this partition. 
The needed terms of the $n$-th order response are of the form $P \mu(t_n)...\mu(t_1) P\rho(t_0)P \mu(t_1)...\mu(t_n)P$ which involve calculating terms of the form $P \mu(t_n)...\mu(t_1) P$.
 We only need to calculate explicitly the first two terms (and the rest follow straightforwardly). Recognizing that $\mu = P \mu P$ (i.e. that $\mu$ is entirely contained in $P$):
\begin{equation}
\begin{split}
    P\mu(t_1)P &= Pe^{iH_0t_1}P\mu Pe^{-iH_0t_1}P \\
    P\mu(t_2) \mu(t_1) P &= Pe^{iH_0t_2}P\mu Pe^{-iH_0t_2}(P+Q)e^{iH_0t_1}P\mu Pe^{-iH_0t_1}P 
\end{split}
\label{eq:terms_mu}
\end{equation}
 The required evolution operator $e^{-iH_0t}$ can more easily be calculated in its resolvent form:
\begin{equation}
    e^{-iH_0t} = \frac{-1}{2\pi i} \oint dz G(z)e^{-izt}
\end{equation}
where $G(z)=[z-H_0]^{-1}$, then for the terms appearing in Eq. \eqref{eq:terms_mu}:
\begin{equation}
\begin{split}
    Pe^{-iH_0t}P &= \frac{-1}{2\pi i}\oint dz PG(z)Pe^{-izt} \\
    Pe^{-iH_0t}Qe^{-iH_0t_1}P &= \frac{-1}{4\pi^2}\oint dz \oint dz' PG(z)Qe^{-izt} QG(z')Pe^{-iz't_1}  \\
  \end{split}
\end{equation}
The resolvent approach allows us to calculate $PG(z)P$ and $PG(z)Q$ of the Hamiltonian in Eq. \eqref{eq:Hamiltonian_diagonalized} using Lippman-Schwinger series. As we have done before,  \cite{FinkelsteinShapiro2020}, we express all of the operators in terms of $\Heff(z) = P\Heff(z)P = PH_0P + PH_0QG_0(z)QH_0P$ where $PG(z)P = [z-\Heff(z)]^{-1}$. In principle, all of the processes that can occur inside the band structure such as electron-electron scattering, electron-phonon scattering appear in the operators that describe the metallic band $QG_0(z)Q$.  However, they are difficult to solve explicitly and go beyond the scope of this work, where we will constrain ourselves to the nonlinear response functions at the delay time between pump and probe $T=0$. We thus neglect any scattering between two states $k$ and $k'$ of the metal. This means that $QG_0Q = [z-QH_0Q]^{-1}$ is diagonal and thus easily invertible. 
To first order, we can make an approximation regarding the energy dependence of the states in partition $Q$ referred to as the wideband approximation that involves an infinite flat continuum with energy independent couplings and a linear dispersion \cite{FinkelsteinShapiro2018}.  
Within these approximations, terms containing $PG(z)Q G(z')P$ all vanish, and $\Heff(z)$ becomes $z-$independent. We can then express the optical response exclusively in terms of the effective operators in the bright partition:
\begin{equation}
    \mu(t) = P\mu(t)P = Pe^{iH_0t}P\mu Pe^{-iH_0t}P = e^{i\Heff^{\dagger} t} \mu e^{-i\Heff t}
\end{equation}

In general, the effective operator $\Heff(z)$ is nonlinear in that it depends on the frequency parameter $z$. The physical meaning of this is that the particle density can transfer from the $P$ partition to the $Q$ partition (and back), which can play a role in the dynamics. This goes beyond the scope of the article, although in the description in the wideband approximation will be valid at short enough times. 
Before understanding the spectral signatures depending on the non-Hermiticity, we isolate the important regimes of Eq. \eqref{eq:Heff} which can correspond to different symmetries of the eigenvalues. 
As mentioned, non-Hermiticity can arise in a number of settings, and is usually connected with a manifold of states that are not explicitly described. The physical effect of the imaginary part of the energy is to destroy (when negative) or create (when positive) particle density. The linear response of a system is not sensitive to what happens to the particle density that the non-Hermitian part of the Hamiltonian destroys. The fate of the particle density only becomes important in nonlinear response (most easily accessible in time-resolved experiments) when the particle density that leaves the system can come back into it. This can be captured exactly by Eq. 6,7, provided that the effective operators can be calculated exactly.
\newline

\begin{figure}
    \centering
    \includegraphics[width=0.8\textwidth]{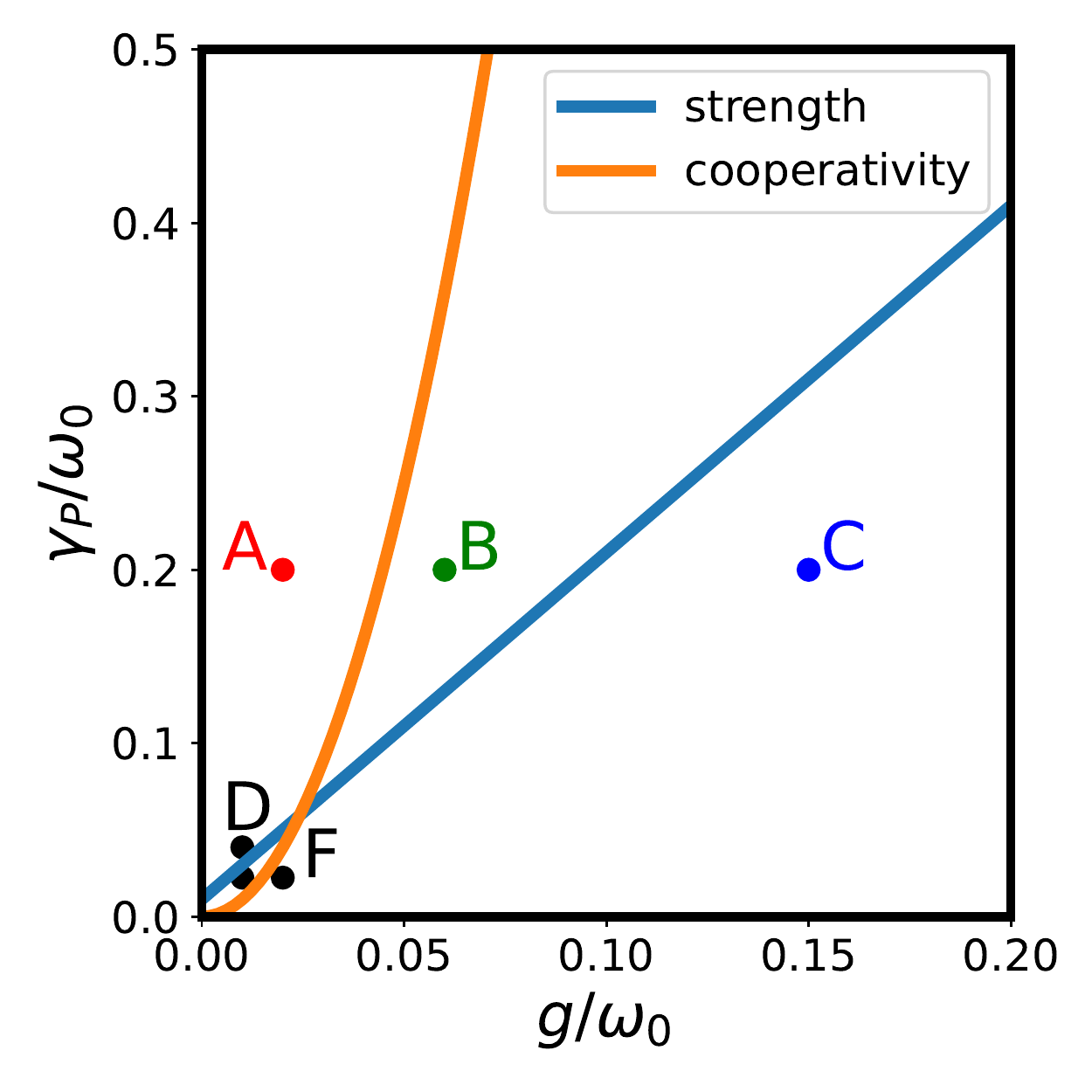}
    \caption{Main regime for a plexciton systems with isoenergetic transitions $\omega_J=\omega_P=\omega_0$. We show in orange and blue the curves for $S=1$ and $C=1$ which separate the different regimes. The parameters of coupling strength $g$ and plasmon dissipation strength $\gamma_p$ are normalized to the molecular transition frequency $\omega_0$. We choose three representative points $A$, $B$ and $C$ to simulate the linear and nonlinear optical response in the main text. The points $D$, $E$ (unlabeled point) and $F$ are explored in the Appendix to characterize the region where $S>1$ but $C<1$) (which is obtained for point $E$, while $D$ and $F$ are provided as references).}
    \label{fig:regimes}
\end{figure}

\textbf{Phenomenology in different coupling strength regimes.} The behavior of polaritonic states can be categorized in regimes according to the strength of the light-matter coupling, the magnitude of the dephasing rates and detuning \cite{Torma2015}. We limit ourselves to coupling strengths for which the rotating-wave approximation is valid. 
For isoenergetic plasmonic and molecular transitions $(\omega_P=\omega_J)$, the non-Hermitian Hamiltonian (see Eq. \eqref{eq:Hamiltonian}) has an exceptional point at $S = 2g/(\gamma_P-\gamma_J) = 1$ \cite{Rodriguez2016}, taken as a boundary between the weak ($S<1$) and strong ($S>1$) coupling regimes. 
An additional dimensionless parameter, the cooperativity $C=g^2/\gamma_P\gamma_J$ has been proposed to separate the region of weak coupling $C<1$ and the region where interference processes are present $C>1$ \cite{Pelton2019,Westmoreland2019}. We will systematically analyze the first and third-order optical response above and below the exceptional point with six parameter families cases marked in Figure \ref{fig:regimes}.a.
Point A (red) corresponds to the weak coupling regime for $C=0.20$ and $S=0.21$, Point B (green) corresponds to the interference regime with $C=1.8$ and $S=0.63$ and Point C (blue) corresponds to the strong coupling with $C=11.25$ and $S=1.58$. The points $D$ ($C=0.25$,$S=0.67$), $E$ ($C=0.44$,$S=1.6$) and $F$ ($C=1.78$,$S=3.2$) are meant to explore a small region where we can have the condition $S>1$ but $C<1$ and is outlined in Appendix C. In all cases the molecular dephasing is set at $\gamma_J/\omega_0=0.01$. As mentioned previously, we consider two limiting cases where the plasmonic transition couples predominantly to the far field (i.e. $\mu_J=0$) and where the molecular transition couples to the far field ($\mu_P=0$). \newline

\begin{figure}
    \centering
    \includegraphics[width=1.0\textwidth]{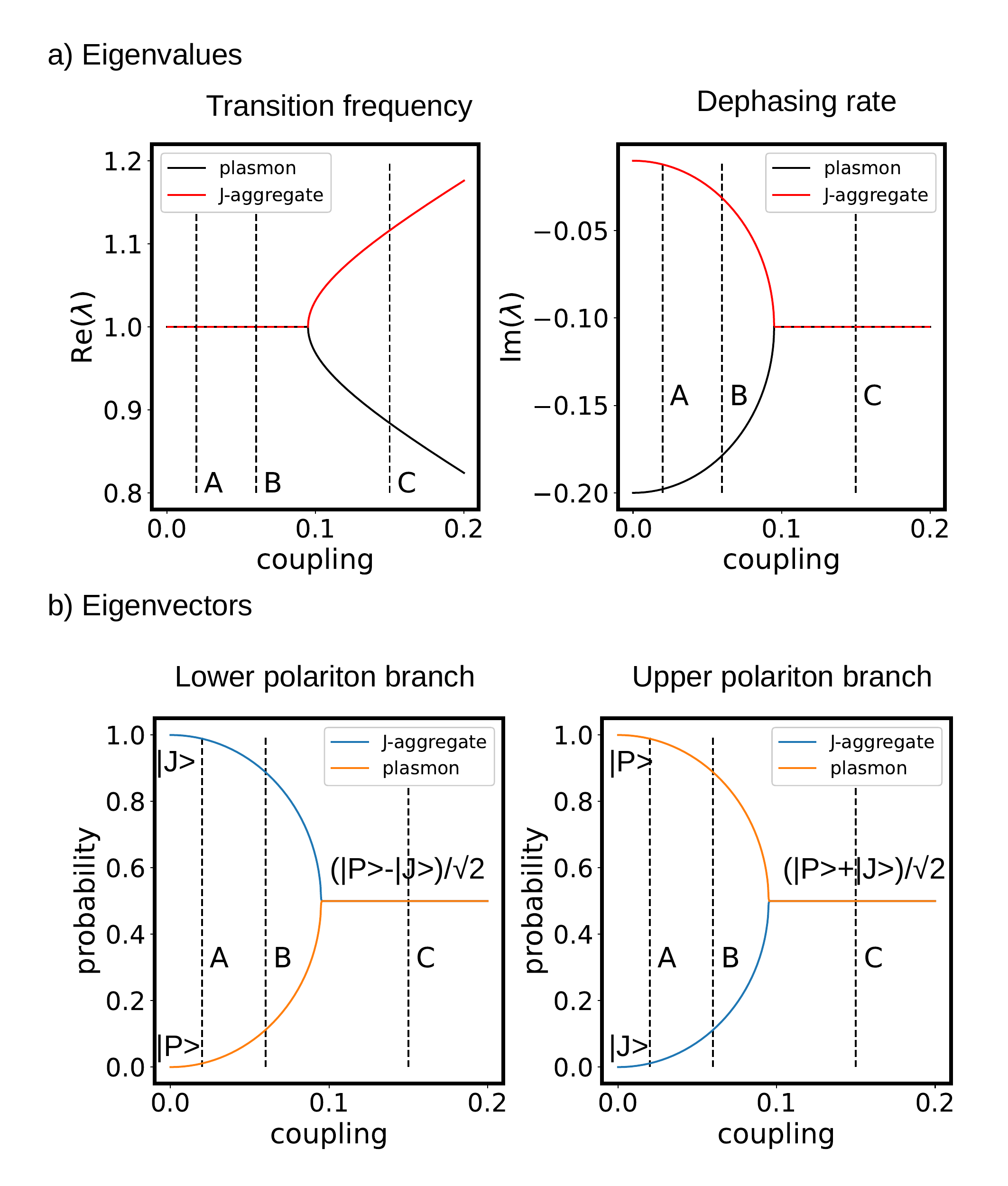}
    \caption{a) Eigenvalues and b) eigenvectors of the non-Hermitian Hamiltonian. Vertical dashed lines show the dipolar coupling strength corresponding to the points A, B and C of Figure \ref{fig:regimes}}
    \label{fig:eigenvalues_eigenvectors}
\end{figure}

As has been pointed out, for example in \cite{Rodriguez2016}, the behavior is different on either side of the exceptional point. We make this explicit by plotting the eigenvalues of Hamiltonian \eqref{eq:effective_Hamiltonian} and eigenvectors as a function of the coupling strength (Figure \ref{fig:eigenvalues_eigenvectors}).
For $S<1$ the real part of the eigenvalues (i.e. the transition frequencies) are equal while the imaginary part (i.e. the dissipation) are not. The two polariton branches are either predominantly plasmonic or predominantly molecular and the distinction is carried by the dephasing rate of the branch (imaginary part of the eigenvalue). At the exceptional point both eigenvalues and eigenvectors coalesce and become indistinguishable. For $S>1$, the real part of the eigenvalues begins to split (i.e. Rabi splitting) while the imaginary part is now identical for both branches. The two polariton branches are now (and for all stronger couplings) an equal mixture of plasmon and exciton. 
We analyze in more detail the first and third order optical response of the non-Hermitian Hamiltonian from Eq.~\eqref{eq:Hamiltonian} using the double-sided Feynman diagram formalism \cite{FinkelsteinShapiro2020} in order to analyze the expected spectral signatures. \newline

\section{Simulations}

We consider the n-th order response of the density matrix using Eq. \eqref{eq:rho_n} and calculate the signal of two-dimensional electronic spectroscopy and linear absorption in the impulsive limit \cite{Mukamel1995}. \newline

\textbf{Linear absorption.} Linear optical absorption is a routine measurement for nanoparticles and can yield a wealth of information at a relatively inexpensive cost, compared to more advanced spectroscopies. In the case of plexcitons, it can reveal the coupling strength, relative magnitude of transition dipole moments and dephasing rates of each individual component. The signal can be obtained from Eq. \eqref{eq:local_absorption}  \cite{Finkelstein2021}: 
\begin{equation}
\begin{split}
S^{(1)} &\propto Re \left(\int dte^{i\omega_t t}\sum_{a=\pm}{\mu}_{a}^2e^{-i{\omega}_{a}t} \right) \\
& = \text{Re}\left( \sum_{a=\pm} \frac{-\mu_{a}^2}{i(\omega_t-\text{Re}(\omega_{a}))+\text{Im}(\omega_a)} \right) \\
& =  \sum_{a=\pm} \frac{-\text{Re}(\mu_{a}^2)\text{Im}(\omega_a)-\text{Im}(\mu_{a}^2)(\omega_t - \text{Re}(\omega_{a}) )}{(\omega_t-\text{Re}(\omega_{a}))^2+(\text{Im}(\omega_a))^2} 
\end{split}
\label{eq:local_absorption}
\end{equation}
where $\ket{+},\ket{-}$ are the eigenstates of the non-Hermitian Hamiltonian (Eq. \eqref{eq:effective_Hamiltonian}) (or upper and lower polariton branches). 

\begin{widetext}

\begin{figure}[h]
    \centering
    \includegraphics[width = 1.0\textwidth]{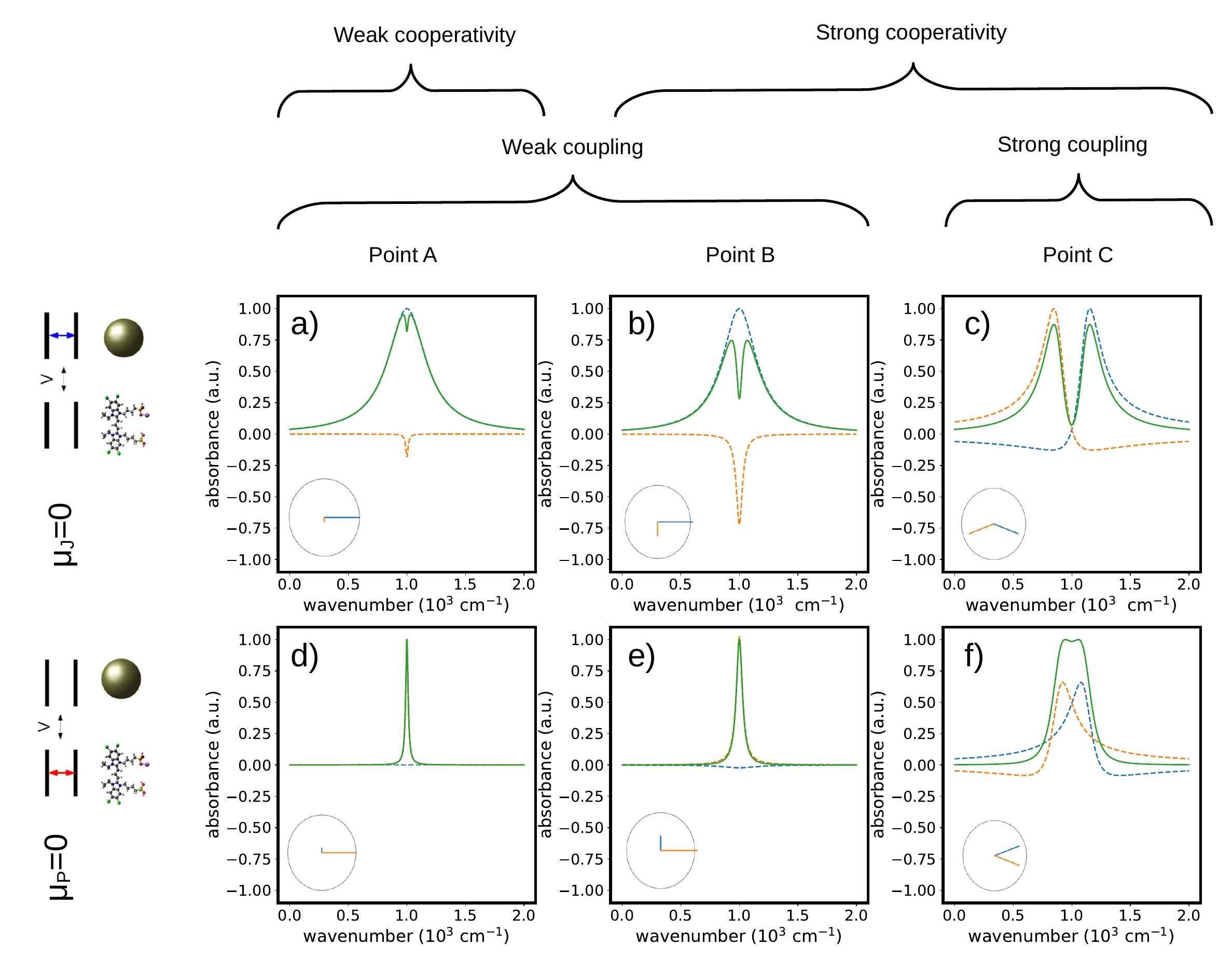}
    \caption{Decomposition of the absorption spectra into polariton branches contributions (dotted yellow and dotted blue give the total contribution for the total absorption shown in green). We show the cases for $\mu_j=0$ (a,b,c) and $\mu_p=0$ (d,e,f). The transition dipole moments to each branch are displayed in the complex plane. We illustrate the regimes of weak cooperativity and weak coupling (a,d), strong cooperativity and weak coupling (b,e) and strong cooperativity and strong coupling (c,f).}
    \label{fig:decomposition}
\end{figure}

\end{widetext}

We can decompose the absorption spectra into each polariton contribution, i.e. plot each term of the sum in Eq. \eqref{eq:local_absorption} separately. In Figure \ref{fig:decomposition}, we show the total absorption (green, solid) decomposed into the upper (blue, dotted) and lower (yellow, dashed) polaritons, for different values of the coupling corresponding to points A, B and C of Figure \ref{fig:regimes}. We consider two limiting cases, one where the plasmon predominantly couples to the far field (top row), and one where the J-aggregate predominantly couples to the far field (bottom row). The Euler plane of the insets show the transition dipole moments for each polariton branch, $\mu_{-},\mu_{+}$. 
We first consider the case where the plasmon couples to the far field (top row). We can see that for all parameters there is the appearance of a peak splitting, however below the exceptional point it arises from the destructive interference of a broad transition and a narrow transition while above the exceptional point it arises from a Rabi splitting.  
Below the exceptional point (A and B), $\mu_{+}$ is purely real while $\mu_{-}$ is purely imaginary. The branch with the purely imaginary transition dipole moment will have a negative contribution to the total absorption and cause an interference dip.
We note that the individual contributions of the branches are not physical by themselves, so that a negative contribution does not mean a stimulated emission but rather intuitively describes the interference process.
The dip is clearly seen for spectra \textbf{a)} and \textbf{b)} where the yellow (dotted) line  cancels the plasmonic absorptive contribution. This results in a spectrum that seems to have  two apparently separate peaks well below the condition for Rabi splitting. However, while the spectra might look similar above and below the exceptional point, the nature of the excitation and consequently the excited state dynamics are expected to be very different. 
As the coupling increases, the transition dipole moments are no longer purely real or purely imaginary but have both real and imaginary components (see the Euler plane insets of Figure \ref{fig:decomposition}). The contribution of each branch now amounts to two true peaks separated by Rabi splitting. While qualitatively the final spectra look similar, the nature of the resonances is not. For example, exciting on the right or left of the dip in the interference regime excites the same homogeneously broadened excited state while in the Rabi splitting regime exciting on the right or left addresses a different excited state (upper polariton or lower polariton). This has been recognized theoretically and experimentally in the fluorescence signatures of quantum dots coupled to plasmons \cite{Leng2018}. 
The case where the J-aggregate couples predominantly to the far field is strikingly different. Below the exceptional point we see no evidence of peak splitting since the negative contribution is too broad. Above the exceptional point we see the expected Rabi splitting although with less contrast than when the plasmon couples to the far-field (Figure \ref{fig:decomposition} d,e).
The interference dip is only observed if it is the plasmon that couples predominantly to the far-field (as is usually the case). The depth of the minimum is modulated by the molecular dephasing rate, and the degree to which the molecular aggregate also couples to the far-field. 
The set of points $D$, $E$ and $F$ show that having the condition $S>1$ is not enough to observe a peak splitting. In the region where $S>1$ but $C<1$ the system where the molecular aggregate couples more strongly to the field does not show a Rabi splitting (see Appendix E, Figure 10). 
While the features of destructive interference are more easily identified, there will also be regions of constructive interference away from the dip.
Thus, both the cooperativity and strength parameters are important for classifying the features of the spectra.
The difference between interference and Rabi splitting has also been observed in simulations with disordered molecules \cite{Engelhardt2022}. There, the narrower cavity absorption generates an interference pattern in the heterogeneously broadened molecular absorption while in the case presented here the narrow molecular absortpion generates an interference pattern in the homogeneously broadened plasmonic absorption. 
 \newline

\begin{figure}
    \centering
    \includegraphics[width = 0.8\textwidth]{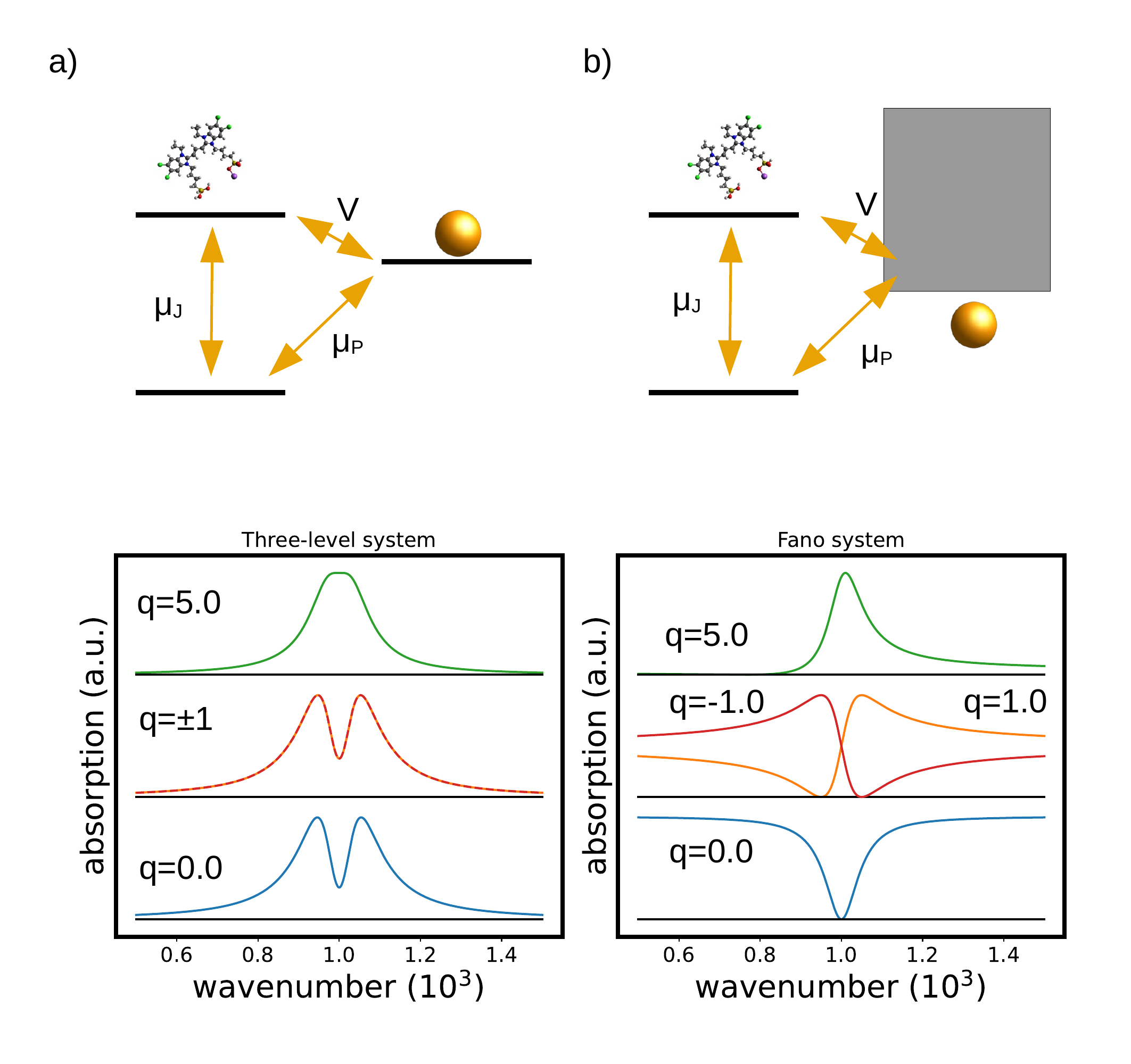}
    \caption{Role of the magnitude of the transition dipole moment of the discrete system $\mu_J$ for a three-level system (a) and a Fano system (b). Traces have been ofset by 1.2 for clarity. The characteristic asymmetry of the Fano model is clearly missing from the three-level system.}
    \label{fig:comparison_Fano}
\end{figure}

\textbf{Fano interferences vs. electromagnetically induced transparency.} It is appropriate to clarify an important point concerning the interference process. This dip is often referred to either as a Fano interference, or an electromagnetically induced transparency (EIT), and both processes denote different physics and spectral signatures. 
The simplest system where EIT is observed is a three-level $\Lambda$ system consisting of a ground state manifold with two levels, and one discrete excited state. Its spectral signature is a frequency region of suppressed absorption. The Fano interference appears in a structure akin to the $\Lambda$ system where one of the levels becomes a continuous manifold of levels and the Fano lineshape is characterized by a distinctive asymmetric lineshape which also includes a frequency region of suppressed absorption \cite{Fano1961}. 
The plexciton Hamiltonian is more closely related to a $\Lambda$ system where instead of having two light-fields coupling each ground state with the excited state, we have an external light field coupling the ground state with the plasmon excitation, and radiative coupling connecting the excited plasmon with the excited J-aggregate (Figure \ref{fig:comparison_Fano}.a).

Figure \ref{fig:comparison_Fano} shows the absorption spectrum for a $\Lambda$ system and a Fano system as a function of the asymmetry parameter which is related to the ratio of the different coupling elements \cite{Fano1961}. For Fano, the asymmetry parameter is $q = \frac{\mu_J}{\pi \mu_P g}$. We choose the same definition for an effective asymmetry parameter for the $\Lambda$ system although its physical meaning is not exactly the same \cite{FinkelsteinShapiro2020}. 
For the Fano model, the limiting cases $\mu_J=0$ ($q=0$) and $\mu_J \to \infty$ ($q\to \infty$) correspond to anti-Lorentzian and Lorentzian lineshapes, respectively. For finite values of $\mu_J$ (or $q$), the Fano system shows distinctive asymmetries that reverse sign as the sign of $q$ is reversed (middle panel), while shifting the condition of destructive interference to the left or right of the zero detuning condition. 
The $\Lambda$ system (Figure \ref{fig:comparison_Fano}.a) shows a similar behavior for the limiting cases $q=0$ and $q=\infty$, if we allow that far away from the resonance condition the absorbance vanishes instead of going to a finite value as in the Fano model. However, for finite values of the effective asymmetry parameter, and unlike the Fano model, we do not find an asymmetry. For finite $q$, the contrast at the point of destructive interference is reduced and the plots are identical for $q$ and $-q$. This point will become important in the analysis of nonlinear signals. We mention that real systems often do show asymmetries, although this is due more to the detuning of the plasmonic transition with respect to the molecular bright transition, and not a true  Fano asymmetry. 

We now turn to a more detailed analysis of the different features that appear in the spectrum of third-order spectroscopies of non-Hermitian Hamiltonians. 

\begin{figure}
\includegraphics[width=0.8\textwidth]{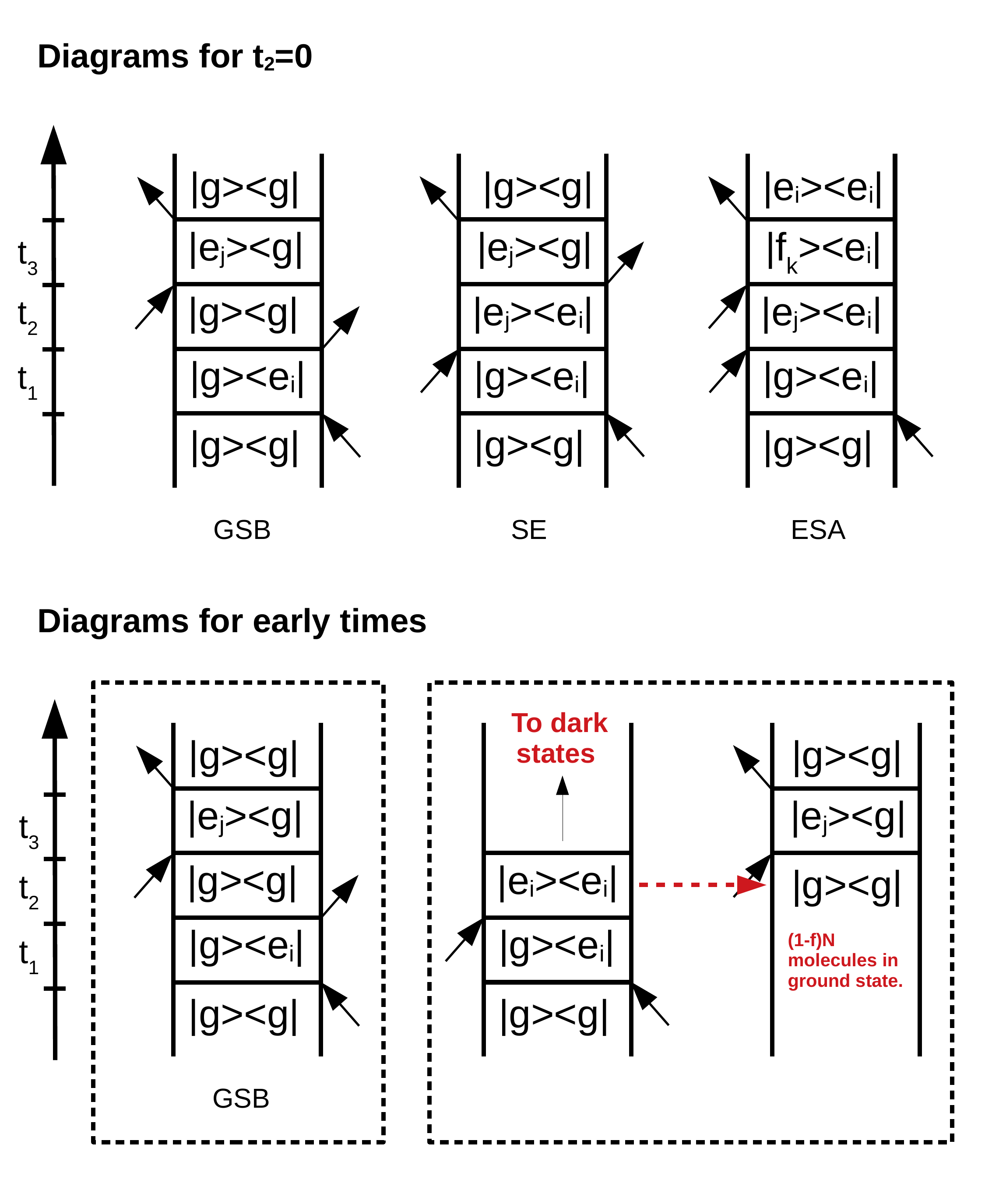}
\caption{Feynman diagrams during the coherent evolution (top) and after dephasing of the plasmon transition. The excited state $e_{i,j}$, $i,j \in LP,UP$ correspond to the polariton branches.}
\label{fig:Feynman_diagrams_2D}
\end{figure}

\textbf{Third order response.} Two-dimensional electronic spectroscopy (2DES) is the most complete third-order spectroscopy. During the experiment, three pulses interact with a sample and an echo is detected (photon-echo 2DES \cite{Jonas2003}) or alternatively four-pulses interact with the sample and an excited state observable is detected, most often fluorescence (action-detected 2D \cite{Tekavec2007,Karki2014}). Of the three characteristic times between the pulses in action detected 2D, the first and last can be Fourier-transformed to obtain two-dimensional plots of the system that show the correlation between excited and detected states at different population times (in photon-echo 2DES the echo is dispersed by a grating and one Fourier transforms along the first delay). By scanning the population time we can reconstruct the dynamics of the excited state. 
The complexity of the signal makes simulations crucial for their understanding, in particular for plasmonic-molecule systems where the signals can be unintuitive \cite{FinkelsteinShapiro2020}. 
The induced polarization can be calculated by the same perturbative approach $P(t)=\text{Tr}(\mu(t) \rho^{(3)}(t))$ where:
\begin{equation}
\begin{split}
    \rho^{(3)} = i^3 \int \int \int dt_1dt_2dt_3 & E(t-t_3) \\
    & \times E(t-t_3-t_{2}) \\
    & \times E(t-t_3-t_2-t_1)\\
    &\langle [\mu(t_2+t_1),[\mu(t_{1}),[\mu(0), \rho(-\infty)] \rangle
\end{split}
\label{eq:rho_3}
\end{equation}
The nested commutator implied by Equation \eqref{eq:rho_3} is now represented by six double-sided Feynman diagrams that are grouped into contributions denoted ground state bleach (GSB), stimulated emission (SE) and excited state absorption (ESA) (see Figure \ref{fig:Feynman_diagrams_2D}). The ground state bleach can be simulated using the fits from the ground state absorption spectrum, while the other two reflect properties of the excited state. The physics captured by the ESA is crucial for the observation of a non-zero nonlinear signal, since it exactly cancels the GSB and SE in linear (harmonic) systems \cite{Mukamel1995}. 

We address in this article the effects of using a non-Hermitian Hamiltonian and not the physics behind the optical nonlinearities in plexcitons or polaritons which have been addressed by others \cite{Ribeiro2018,DelPo2020}. Consequently, we will focus on the GSB contribution and analyze its structure for the different regimes exemplified by points A, B and C of Figure \ref{fig:regimes} and of points D, E and F in the Appendix C.  The other pathways SE and ESA, however, share similar features and the conclusions related to the symmetry of the final signal remain  unchanged (see Figure \ref{fig:GSB_and_ESA} in Appendix D for the real part of the total signal in the case of the Rabi contraction nonlinearity shown in Figure \ref{fig:Feynman_diagrams_2D}). 
In addition, we only consider the case where the plasmon couples predominantly to the far-field so that we can contrast the lineshapes of interferences with that of Rabi splitting. We can write the GSB contribution as

\begin{equation}
\begin{split}
    S^{(3)} & \propto \sum_{i,a=\pm} \frac{\text{Im}(\omega_a)(\text{Im}(\mu_i^2\mu_a^2)(\omega_t-\text{Re}(\omega_i) )+\text{Re}(\mu_i^2\mu_a^2)\text{Im}(\omega_i))}{((\omega_t-\text{Re}(\omega_i) )^2+\text{Im}(\omega_i)^2)((\omega_\tau-\text{Re}(\omega_a) )^2+\text{Im}(\omega_a)^2)}
\end{split}
\label{eq:S3_analytic}
\end{equation}
\begin{figure}
\includegraphics[width=0.8\textwidth]{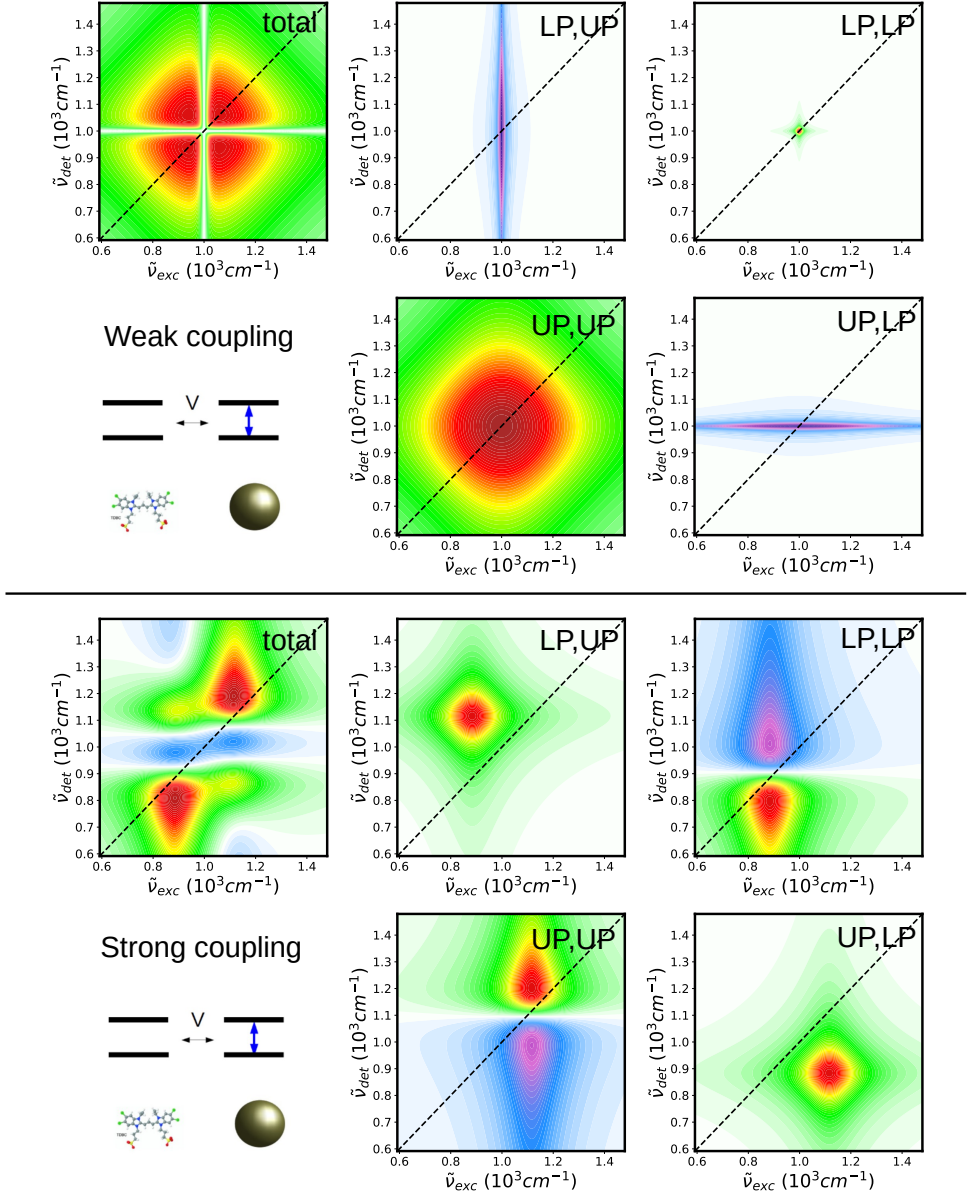}
\caption{Decomposition of the GSB contribution for the case of weak coupling (top) and strong coupling (bottom). Each panel shows the individual contributions corresponding to the different permutations $e_{i},e_{j}=$LP,UP of the diagram in Figure \ref{fig:Feynman_diagrams_2D}}
\label{fig:decomposition_2D}
\end{figure}

Figure \ref{fig:decomposition_2D} shows the spectrum for weak (\ref{fig:decomposition_2D}.a) and strong (\ref{fig:decomposition_2D}.b) coupling along with their decomposition into diagonal elements ((UP,UP) and (LP,LP)) and cross-terms ((UP,LP) and (LP,UP)). For the weak coupling case, the diagonal elements are positive while it is the cross-terms that provide the negative signal and split the main plasmonic resonance into four regions. That the negative features arise from the cross-terms between upper and lower branch is a clearer indication that the dips are a result of an interference process between the plasmonic resonance and the molecular resonance.

The strong coupling regime is qualitatively different. Each of the four contributions now appear at a different position in the spectrum because of the Rabi splitting. 
A notable difference between the GSB signal between the two regimes is that in the interference regime has a $D_4$ fourfold reflection symmetry while the strong coupling regime possesses a $C_2$ rotational symmetry (as we show later when both $\mu_J \neq 0$ and $\mu_P \neq 0$ some symmetries are lost). 

\begin{figure}
    \centering
    \includegraphics[width=0.95\textwidth]{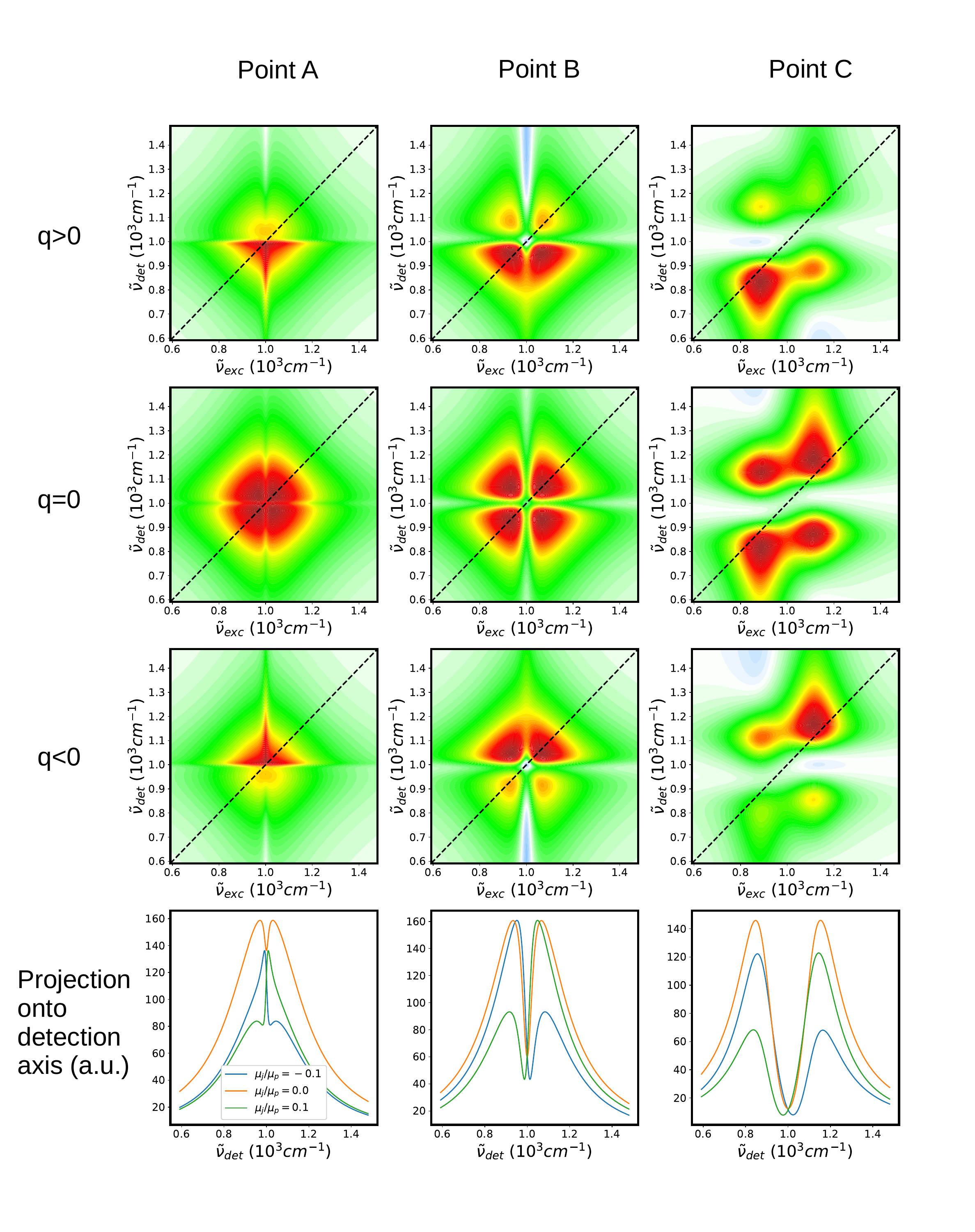}
    \caption{Dependence of the spectral asymmetry along the detection dimension as a function of the ratio $\mu_J/\mu_P=$ -0.2 (top row), 0 (middle row) and 0.2 (bottom row). The coupling strength increases from left to right.}
    \label{fig:muJ_dependence}
\end{figure}

We now analyze the effect of having both $\mu_{P} \neq 0$ and $\mu_{J} \neq 0$. 
Figure \ref{fig:muJ_dependence} shows the GSB for different values of the coupling strength $g$ (columns) and also different values of the ratio $\mu_J/\mu_P$ (rows). Along the columns we observe the same qualitative difference between weak and strong coupling, namely the transition in the symmetry of the lineshape. In this case when $\mu_J \neq 0$, we loose some symmetries. Strikingly, while the linear response for different values of $\mu_J/\mu_P$ does not show an asymmetry as the effective asymmetry parameter changes, the third-order response shows a clear asymmetry along the detection dimension. Projections along the excitation dimension (lower row of Figure \ref{fig:muJ_dependence}) show that this asymmetry will be visible in transient absorption. This suggests that differences in the linear spectrum and the 2D projection onto the detection dimension can reveal information on the Hamiltonian. Asymmetries in the linear response will denote different transition frequencies between the plasmonic and molecular parts, while additional asymmetries in the transient absorption will indicate finite values of $\mu_J$. 

\section{Discussion}

\textbf{General comment on non-Hermitian Hamiltonians}.
The description of physical phenomena using non-Hermitian Hamiltonians is pervasive and extends far beyond the physical systems described here \cite{Bender2007}. These can exist in metamaterials, engineered Floquet states and gain/loss media in general. The treatment described in the previous section applies to all of these systems as long as they have a non-zero nonlinear response. Our findings suggest that it is interesting to consider two-dimensional spectroscopy (or more generally higher orders) as tools that can  distinguish between different symmetries of the eigenvalues by encoding them in symmetries of the signal. \newline
Rigorously, a Lindblad operator should be used instead of a non-Hermitian Hamiltonian. However, the quantum jump operator that restores the particle density back to the ground state can be encoded by a judicious choice of Feynman diagrams. 

\textbf{Comparison to other models in the literature}. There exist many models to describe the response of emitters coupled to plasmonic resonances which can fit absorption spectroscopy, scattering or photoluminescence data well. 
For example, the coupled oscillators model can explain the regimes of interference and Rabi splitting and reproduce the measured lineshapes. However, the formalism is not apt to describe nonlinear spectroscopy because of the impossibility of isolating the signals to a certain order in the field. 
\textit{Ab initio} approaches involving the solution of Maxwell's equations around a metallic nanoparticle result in Fano-like equations with slightly modified detunings, or  in non-Hermitian Hamiltonians for molecules coupled to plasmonic resonances. These expressions are sometimes reflective of scattered signals and not absorptive signals \cite{Faucheaux2014,Gallinet2010,Pelton2019}, as such they are suitable for linear response but not nonlinear signals. 
The use of non-Hermitian Hamiltonians as justified above solves this problem and allows the use of the response function formalism and double-sided Feynman diagrams to describe interference phenomena, and in particular to suggest its use for discerning between interference processes and Rabi splitting. We mention that it is not the only option and previous works that compare scattering and photoluminescence (PL) show a marked difference in the Rabi splitting with significant reductions in the PL spectra compared to the scattered light \cite{Pelton2019}. For regions of interference, the peak splitting entirely disappears from the PL spectrum while it is visible in the scattered spectrum \cite{Leng2018,Gupta2021}, resulting in a diagnostic tool which can be applied in the case of few emitters coupled to a plasmonic nanoparticle. \newline

\textbf{Symmetries of the eigenvalues and symmetries of the signal.} One of the findings of the paper is that the third-order signal is qualitatively sensitive to changes in the symmetry of the eigenvalues, unlike the first order signal. 
The symmetry of the linear and nonlinear signal with respect to reflection along $\omega_\tau = \omega_0$ and $\omega_t = \omega_0$ can be analytically obtained by the symmetries in the eigenvalues and eigenvectors below and above the exceptional point. We define the detunings with respect to the transition frequencies as $\delta_i = \omega_i-\omega_0$, and the reflection operations that take $\delta_\tau \to -\delta_\tau$ and $\delta_t \to -\delta_t$ as $\mathcal{O}_{\tau}()$ and $\mathcal{O}_{t}()$, respectively. Below the exceptional point, we have the following properties of the eigenvalues and transition dipole moments: $\delta_{+}=\delta_{-} = 0$, $\mu_{+}$ is purely real and $\mu_{-}$ purely imaginary. 
It is straightforward to verify that $\mathcal{O}_{t}(S^{(1)})=S^{(1)}$. In the strong coupling limit, we have that  $\mu_{-}^2=(\mu_{+}^2)^*$, so that it is also verified that because $\delta_{-}=-\delta_{+}$, $\mathcal{O}_{t}(S^{(1)})=S^{(1)}$ and the signal is symmetric with respect to the central transition frequency $\omega_0$. 

In a similar fashion we can calculate the symmetries of the third order signal described by equation \eqref{eq:S3_analytic} using equations \eqref{eq:eigenvalues} and \eqref{eq:eigenvectors}. Below the exceptional point, we have that $\mathcal{O}_{t}(S^{(3)})=\mathcal{O}_{\tau}(S^{(3)})=S^{(3)}$, and above it we recover that $\mathcal{O}_{t}(\mathcal{O}_{\tau}(S^{(3)}))=S^{(3)}$. 
This work raises the possibility of using the symmetry of the lineshape of nonlinear spectroscopy to study the symmetry of the eigenvalues of the Hamiltonian. 
In real systems we expect to have deviations from these idealized Hamiltonian, however we do expect the symmetries of the spectra to be recognizable.

\section{Conclusion}

We have derived a non-Hermitian Hamiltonian for the calculation of the linear and nonlinear optical response. Decomposing the Hamiltonian into bright and dark partitions and expressing the linear and nonlinear optical response in terms of effective operators provides a new framework for systems with large manifolds of dark states. 
An analysis of the linear optical absorption reveals that the exceptional point separating weak from strong coupling regime is clearly illustrated when decomposing the total signal into individual Feynman diagram contributions. Below the exceptional point interference effects are indicated by negative contributions to the absorbance while the Rabi splitting appears above the exceptional point. 
The decomposition of the nonlinear signal also provides additional insight. We find that interference and Rabi splitting regimes have different symmetries in the two-dimensional maps, an indication that cannot be obtained from linear absorption alone. 
The connections outlined in this article between the symmetry of the spectral signatures and the nonlinear response can open new ways of thinking for classifying symmetries in the eigenvalue/exciton structure of complex materials.

\clearpage

\bibliography{ebbesen,Fano,from_second_review,J-aggregates,plexcitons,last_refs,extra_plexciton}

\section{Appendix A. Derivation of a Non-Hermitian Hamiltonian for plexcitons}

\setcounter{equation}{0}
\renewcommand{\theequation}{A.\arabic{equation}}

In this Appendix, we derive the effective Hamiltonian corresponding to a plexciton. We consider similar parameters to a physical sample we have measured  previously and base our approximations on its values (see \cite{Finkelstein2021} for the parameters). 

\begin{equation}
\begin{split}
H & = H_{molecule} + H_{metal} \\
&+ H_{metal-molecule} + H_{\text{field}} + H_{\text{fluctuations}}(t)  \\
H_{molecule} & = \sum_i^N\omega_0 a^{\dagger}_{J,i} a_{J,i}  + \sum_{i,j \in n.n.} K a^{\dagger}_{J,i}a_{J,j} \\
H_{\text{metal}}&= \omega_P a^{\dagger}_P a_P + \int dk \omega_k a^{\dagger}_k a_k \\
& + \int dk (V_ka^{\dagger}_k a_P + V_k^*a^{\dagger}_P a_k) \\
H_{\text{metal-molecule}} & = \sum_i g_0(a^{\dagger}_{J,i} a_P + a^{\dagger}_P a_{J,i}) \\
H_{\text{Field}} & = E(t)\sum_i \mu_0 (a^{\dagger}_{J,i} + a_{J,i}) + \mu_P (a^{\dagger}_P + a_P) \\
H_{\text{fluctuation}}(t) &= \sum_i \Delta \omega_{ii}(t) a^{\dagger}_{J,i} a_{J,i}
\end{split}
\label{eq:Hamiltonian}
\end{equation}
where $a^{\dagger}_{J_i}$ ($a_{J_i}$) are the creation (annihilation) operators for the $i$-th molecule and K is the nearest neighbor coupling between molecules, $a^{\dagger}_P$ ($a_P$) are the creation (annihilation) operators for the plasmon mode, $a^{\dagger}_k$ ($a_k$) those of the manifold of continuum states in the metal band structure with momentum $k$, $V_k$ the plasmon-electron coupling responsible for Landau decay. $g_0$ the dipolar coupling between the molecular transitions and the plasmon mode, which we assume to be identical for all molecules. $H_{\text{Field}}$ is the light-matter interaction with an external field $E$, where $\mu_0$ is the transition dipole moment of a molecule and $\mu_P$ that of the plasmon. $H_\text{fluctuation}(t)$ represents the fluctuations induced by the bath, where we have only considered the modulation of the transition frequency of the $i$-th molecule with amplitude $\Delta \omega_{ii}(t)$. The Hamiltonian is depicted in Figure \ref{fig:Hamiltonian}.a.

There are two main line-broadening mechanisms. On one hand, the optical coherence of the molecular component dephases with fluctuations of the transition frequency (with a time constant of 40 fs \cite{Finkelstein2021} for TDBC, which is much faster than the decay induced dephasing into molecular dark states). To capture this dephasing mechanism, it is enough to consider the stochastic fluctuation of the transition frequency of the molecules. The plasmon coherence, on the other hand, dephases via Landau damping whereby the plasmon decays into neutral hot electron-hole states inside the metal. For this process, we can consider the coupling of the plasmon transition to a a continuum of electron-hole states via a plasmon-electron coupling element \cite{RomanCastellanos2019}. These two are very different mechanisms. The lifetime induced broadening due to decay of the plasmon into hot electron-hole states is not a thermal process and has negligible temperature dependence, in stark contrast to line broadening in molecules induced by fluctuations of the bath which have strong temperature dependence and can be significantly reduced at low temperatures.

With Eq. \eqref{eq:Hamiltonian}, we model the molecular aggregate as $N$ two-level systems which are coupled to their nearest neighbour with strength $K$. The plasmon is represented by a boson coupled to each two-level system with a coupling strength $g_0$. The plasmon excitation can decay into hot electron-hole pairs that form a continuum we label by its momentum $k$. The decay is mediated by the plasmon-electron interaction $V_k$ \cite{RomanCastellanos2019}. Both the molecules and the plasmon couple to the far-field with transition dipole moments $\mu_0$ and $\mu_P$, respectively (see Figure \ref{fig:Hamiltonian}). These couplings to the light field are analyzed in the main text for the two limiting cases when $\mu_0=0$ and when $\mu_P=0$.  
%

There are three energy scales of the problem: the coupling strength $g=g_0\sqrt{N}$, the nearest neighbor coupling $K$ and the transition energy fluctuation amplitudes $\Delta \omega_{ii}$. To derive the current effective Hamiltonian we assume the limit $\Delta \omega_{ii} \gg g, K$ which is appropriate for the study of interference effects. For simplicity we also assume here the limit of vanishing intermolecular coupling $K$.  
The approach is to diagonalize the Hamiltonian for the molecular component, average over fluctuations over the bath, couple to the plasmonic part and project onto the bright states. 

In the molecular basis, the transition frequency of molecule $i$ is modulated by the bath by an amount $\Delta \omega_{ii}(t)$. 
We transform $H$ from the molecular basis to the molecular exciton basis via the unitary transformation $W$ to obtain a diagonal Hamiltonian $D$:
\begin{equation}
    D  = W H W^{-1}
\end{equation}
where $W$ are the eigenvectors of the Hamiltonian without fluctuations. We index the new eigenstates by $q$ and transform the fluctuations into the molecular exciton basis as well: 
Then:
\begin{equation}
\begin{split}
H_{\text{exciton}} & = H-H_\text{metal}-H_\text{metal-molecule}-H_{field} = \sum_{q=0}^{N} \omega_q a^{\dagger}_q a_q + 
 \sum_{q=0}^{N} \sum_{l=0}^{N} \delta_{ql} (t) a^{\dagger}_q a_l
\end{split}
\label{eq:Ap:molecular_Hamiltonian_diagonalized}
\end{equation}
The $q=0$ is bright and is responsible for coupling to the far field as well as for the coupling to the plasmonic transition. The flucutations in the new basis are given by:
\begin{equation}
    \delta_{ql}(t) = W_{qi} \Delta\omega_{ii} (t) W^{-1}_{il}
\end{equation}
We can neglect the off-diagonal terms $\delta_{ql}$, for $q\neq l$ so that only the diagonal fluctuations of the eigenstates survive \cite{Cho2009}. 
The cumulant expansion of the evolution operator $\langle e^{-i\int dt' H(t')} \rangle_B = e^{-i\langle H\rangle t}$ can be carried out straightforwardly: 
\begin{equation}
    \langle e^{-i\int dt' H(t')} \rangle \approx e^{-iH_m t-\Gamma t}
\end{equation}
where $H_m = H_{exciton}-H_\text{fluctuations}$ and $\Gamma = \sum_q \gamma_q a^{\dagger}_qa_q$ and $\gamma_q=\int dt' e^{i \omega t} \delta_{qq}(t') \delta_{qq}(0)$ is obtained from the two-time correlation function. Assuming the Markovian limit, $\gamma_q$ is real and each excitonic state dephases with a different time constant. We will only concern ourselves with the $q=0$ level corresponding to the bright state. 

The new Hamiltonian is:
\begin{equation}
\begin{split}
H & = H_{\text{bright}} + H_{\text{dark}} + H_{\text{bright-dark}} + H_{\text{Field}} + \langle H_{\text{fluctuation}} \rangle \\
H_{\text{bright}} & = \omega_J a^{\dagger}_J a_J + \omega_P a^{\dagger}_P a_P + \sum_i g(a^{\dagger}_{J} a_P + a^{\dagger}_P a_{J})  \\
H_{\text{dark}} & = \sum_{q=1}^{N} \omega_q a^{\dagger}_q a_q + \int_{-\infty}^{+\infty} dk \omega_k a^{\dagger}_k a_k  \\
H_{\text{bright-dark}} & = \int dk (V_ka^{\dagger}_k a_P + V_k^*a^{\dagger}_P a_k)  \\
H_{\text{Field}} & = E(t) \left( \mu_J (a^{\dagger}_{J} + a_{J}) + \mu_P (a^{\dagger}_P + a_P) \right) \\
\langle H_{\text{fluctuation}} \rangle &= -i \sum_{q=0}^{N}  \gamma_q a^{\dagger}_q a_q
\end{split}
\label{eq:Hamiltonian_diagonalized}
\end{equation}
where $g=\sqrt{N}g_0$
Assuming a tight-binding model with identical site energies and identical nearest neighbor coupling $K$, the dispersion relation for the molecular component is $\omega_q = \omega_0 - \sqrt{K^2(2cos(\frac{2\pi}{N}q)+2)}$ \cite{Petersson2000}. 

The Hamiltonian (Eq. \eqref{eq:Hamiltonian_diagonalized}) can be already used to calculate the optical response of the plexcitons, however, the resulting expression for the signal will contain an explicit description of dark states that do not couple to the far-field and so is inefficient. In order to transform to a fully excitonic picture we must also diagonalize the plasmon-metal Hamiltonian. However, it is much more preferable and transparent to obtain effective operators for the bright transitions. \newline

\textbf{Effective operators.} We have obtained the Hamiltonian after bath averaging in the limit of vanishing dipolar interaction and now proceed to calculate the effective operators for the bright and dark parts. We have \cite{FinkelsteinShapiro2020}:
\begin{equation}
\Heff(z) = PH_0P + PH_0QG_0(z)QH_0P
\end{equation}
where $QG_0(z)Q = [z-QH_0Q]^{-1}$. 
For the evaluation of $\Heff(z)$ we assume the wideband approximation where $\omega_k = \frac{k}{n}$ with $n$ a density of states and $V_k$ is $k-$independent. Then, considering that only the metallic dark states are the only ones to affect significantly the optical transitions, we have
\begin{equation}
    PHQG_0(z)QHP = \int dk \frac{\abs{V_k}^2}{z-k/n} = -i n \pi \abs{V_k}^2 
\end{equation}
We also include here the calculation of terms needed to calculate the most general form of the optical response:

\begin{equation}
    PG(z)Qe^{-izt} QG(z')Pe^{-iz't_1} =  PG(z)P \int dk PH_0Q \frac{Q}{z-k/n} \frac{Q}{z'-k/n} e^{-izt}e^{-iz't_1} QH_0PG(z')P
\end{equation}
which vanishes due to the wideband approximation. 
We then obtain the effective Hamiltonian in $P$
\begin{equation}
\begin{split}
    \Heff &= (\omega_J-i\gamma_J )a^{\dagger}_J a_J + (\omega_P-i\gamma_P) a^{\dagger}_P a_P  \\
    &+ g(a^{\dagger}_J a_P + a^{\dagger}_P a_J) \\
\end{split}
\label{eq:effective_Hamiltonian}
\end{equation}
where $\gamma_P= n\pi \abs{V_k}^2$ and $\gamma_J = \int dt e^{i\omega_J t} \langle \delta_{00}(t)\delta_{00}(0) \rangle$. 
The non-Hermitian model presented accounts for plasmon-electron coupling, as well as fluctuation of the energy levels. It is expected that more complicated relaxation schemes, notably relaxation from upper to lower polariton branch in concert with pure dephasing are not fully described. The effect of these pathways will be explored in future work.

\section{Appendix B: Derivation of non-Hermitian operators of levels coupled to continua}

\setcounter{equation}{0}
\renewcommand{\theequation}{B.\arabic{equation}}

The non-Hermitian Hamiltonian of Eq. \eqref{eq:effective_Hamiltonian} appears only under some limiting conditions of the magnitudes of molecular line broadening induced by fluctuations of the bath, the decay rate into continuum states, and dipolar coupling. More general cases quickly become complicated and untractable analytically \cite{Seibt2013}. However, other extended Hamiltonians reduce to the same expression, the simplest of them being two coupled excited states, each one coupled to its own set of continuum states. In general, for a set of $N$ discrete states coupled to each other and to $M$ continua via couplings $V_i^{(a)}$ (corresponding to the coupling of level $i$ with continuum $a$) gives \cite{FinkelsteinShapiro2018}:
\begin{equation}
\begin{split}
PH_0P&=\sum_{i=1}^N \omega_i \ket{i}\bra{i} + \sum_{\substack{i,j=1 \\ i\neq j}}^N g_{ij} \ket{i}\bra{j} \\
H_{\text{eff}}-PH_0P&=-i\sum_{a=1}^M\sum_{i,j=1}^{N}n^{(a)}\pi V_{i}^{(a)}V_{j}^{(a)}\ket{i}\bra{j} \\
\end{split}
\label{eq:recipe}
\end{equation}
where $V_i^{(a)}$ is the coupling (including radiative coupling) between level $i$ and the continuum $(a)$ and
which reduces to equation \eqref{eq:effective_Hamiltonian} with $\gamma_P = n\pi \abs{V_P^{(1)}}^2$ and $\gamma_J = n \pi \abs{V_J^{(2)}}^2$ where we have labeled the individual continua as (1) and (2) for the plasmon and J-aggregate, respectively. 
Notice that if two levels are coupled to the same continuum we obtain off-diagonal non-Hermitian terms.

\section{Appendix C: Extended parameter space for $S>1$ but $C<1$}

\begin{figure}
    \centering
    \includegraphics[width = 1.0\textwidth]{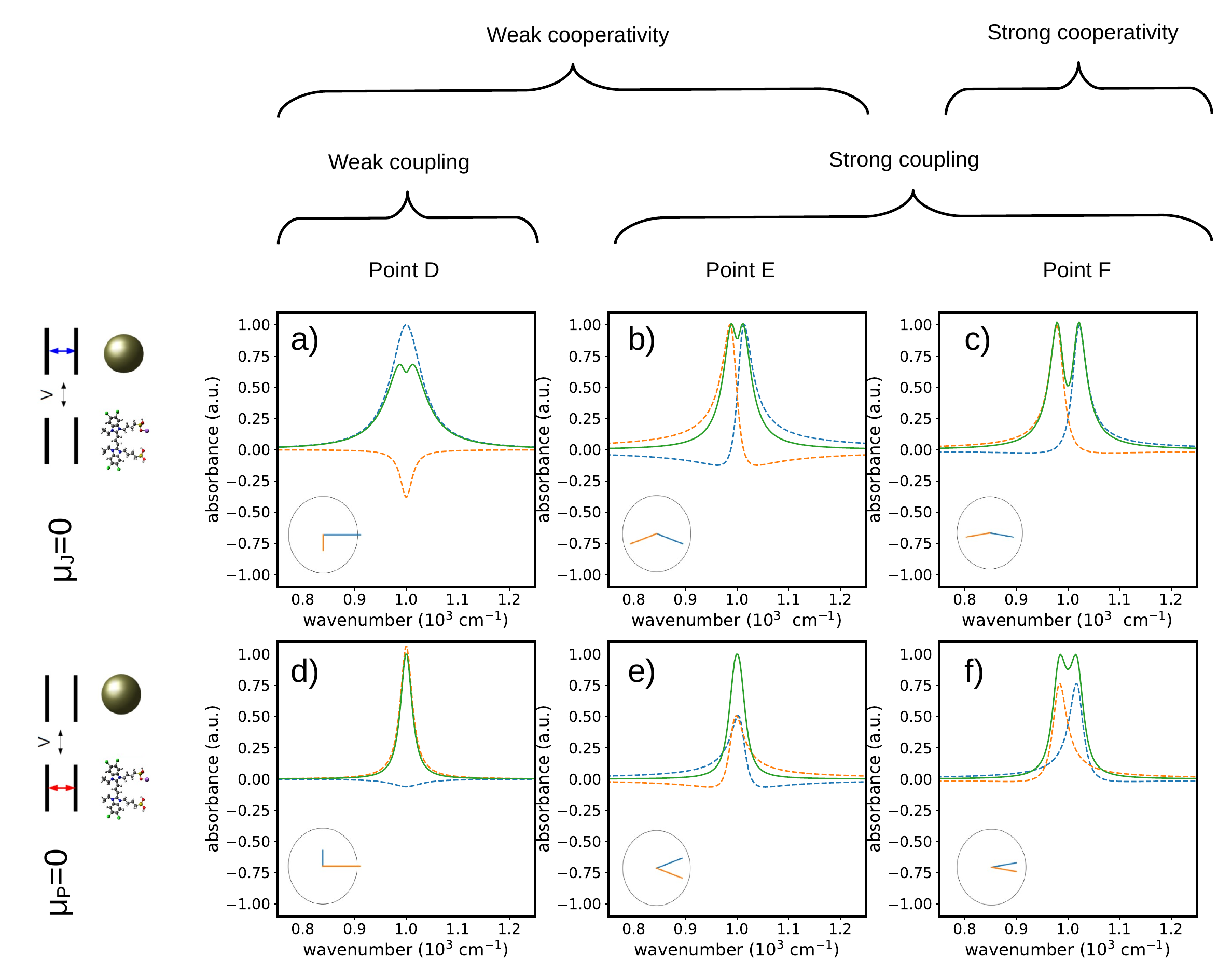}
    \caption{Decomposition of the absorption spectra into polariton branches contributions (dotted yellow and dotted blue give the total contribution for the total absorption shown in green). We show the cases for $\mu_j=0$ (a,b,c) and $\mu_p=0$ (d,e,f), for points D, E and F of Figure \ref{fig:regimes}. We illustrate the regimes of weak cooperativity and weak coupling (a,d), weak cooperativity and strong coupling (b,e) and strong cooperativity and strong coupling (c,f).}
    \label{fig:decomposition_other_points}
\end{figure}

In this Appendix we explore the region marked by points $D$, $E$ and $F$ of Figure \ref{fig:regimes} that are in the strong coupling regime ($S>1$) but not in the weak coupling regime with respect to the interference parameter ($C<1$). Figure \ref{fig:decomposition_other_points} shows that point $E$ shows the presence of Rabi splitting for the case where the plasmon couples to the far field but not for the case where the molecular aggregate couples to the far field. We see here the value of the cooperativity parameter. In the case where the molecular component couples predominantly to the far field, the Rabi splitting is absent even though we are in the strong coupling limit, if we are in the low cooperativity regime. It takes both parameters to be larger than one to observe some the Rabi splitting. 

\section{Appendix D: Inclusion of ESA}

We can simulate the full spectrum by choosing a model for the excited state absorption (ESA). If the excited state absorption is dominated by the molecules remaining in the ground state after the pump pulse pair, we can simulate the full spectrum by the Feynman diagrams shown in Fig. \ref{fig:Feynman_diagrams_2D}. The differences found with experimental measurements can be ascribed to the finite bandwidth of the pulses. 

\begin{figure}
    \centering
    \includegraphics[width=\textwidth]{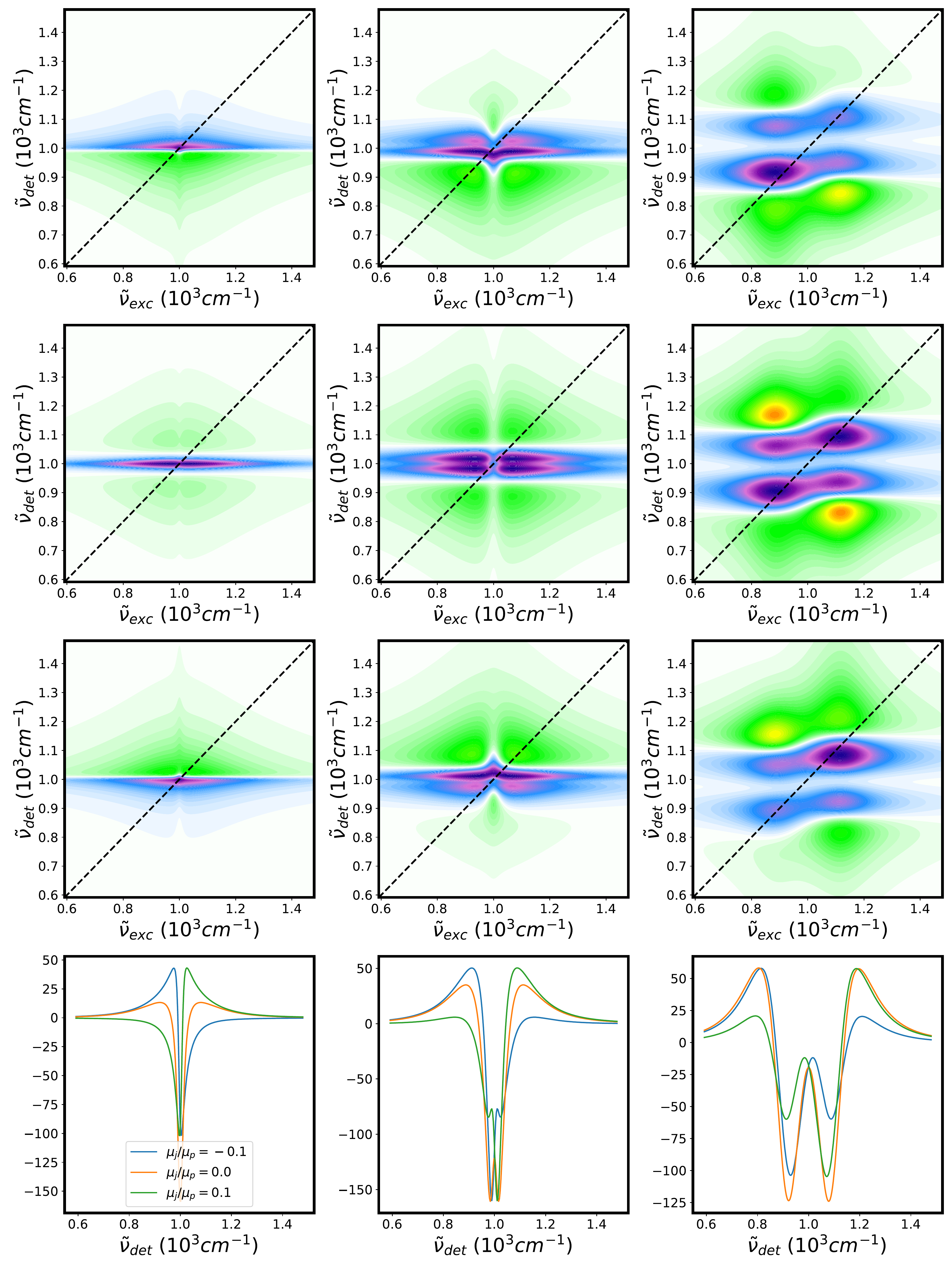}
    \caption{Dependence of the spectral asymmetry for GSB and ESA along the detection dimension as a function of the ratio $\mu_J/\mu_P=$ -0.2 (top row), 0 (middle row) and 0.2 (bottom row). The coupling strength increases from left to right.}
    \label{fig:GSB_and_ESA}
\end{figure}

\end{document}

\section{Notes for me}

We consider a non-Hermitian Hamiltonian with right eigenvectors $v_R$ and left eigenvectors $v_L$. In our Hamiltonian we have that $v_R=v_L$ and normalize them so that $v_R^2=1$.

\end{document}